\documentclass[12pt]{article}

\usepackage{epsfig}

\begin{document}

\title{\bf On a Possible Stationary Point in High-Energy Scattering}

\author{V.A.~Petrov and A.P.~Samokhin \\
Institute for High Energy Physics,\\
``NRC Kurchatov Institute, Protvino'', RF}

\date{}

\maketitle

\begin{abstract}
We discuss a curious observation: at energies from the ISR and up to
the LHC, inclusively, the differential cross-section of elastic
proton-proton scattering remains almost energy-independent at the
transferred momentum $t\approx - 0.21 \mathrm{GeV}^{2}$ at the level
of $\approx 7,5 \mathrm{\ mb}/\mathrm{GeV}^{2}$. The latter value can be
considered as a prediction for $ d\sigma/dt$ at 13 TeV. We also
obtain a lower bound for the forward $ pp $  slope at 13 GeV.
\end{abstract}

\begin{figure}[t]
\centering
\includegraphics[height=9.2cm]{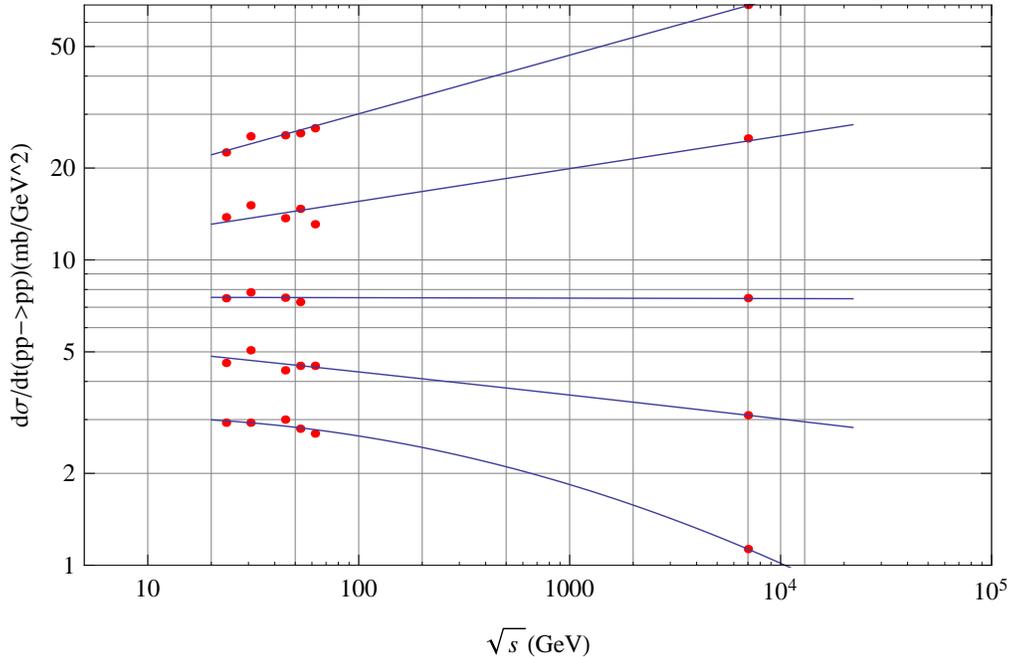}
\caption{Energy evolution of the differential cross-section of
elastic $pp$ scattering at fixed values of transferred momenta. The
values of $-t$ go down as:  $ 0.1$ GeV$^{2}$, $0.15$ GeV$^2$, $0.21$
GeV$^2$, $0.25$ GeV$^2$, $0.3$ GeV$^2$.}
     %\label{}
\end{figure}

\section{Hints at the Stationary Point}
It is well and long ago known that at $ -t  < 0.15$ GeV$^{2}$ the $
pp $ differential cross-section grows with energy. On the contrary,
at $ -t > 0.3$ GeV$^{2}$ they clearly decrease[1]. Recent LHC data
confirm these both  trends.

This circumstance may impose a speculation that there could exist
between these two regions of $ t $ a point $ t_{\ast} $ where
differential cross-section doesn't change with energy at all. The
stationary points of the differential cross-sections were discussed
\textit{en passant} many years ago in Refs. [2],[3] but didn't find
any conceptual development. We, thus, assume that there exists a
fixed (energy independent) point $ t_{\ast}$ where
\begin{center}
$ d [(d\sigma/dt)(s,t_{\ast})]/ds \approx 0 $
\end{center}
at "high enough energies". The thorough inspection of the available
data shows that such a point seems to exist indeed. At Fig.1 the
energy evolution of $ {d\sigma_{pp}}/{dt} $ at various fixed values
of $ t $ is presented.

Fig.1 bears an evidently qualitative character. We believe, however,
that it demonstrates our point quite well. Certainly, the exact
value of the stationary point $ t_{\ast} $,  which we believe at the
moment is near $-0.21$ GeV$^{2}$,  may  slightly change after more
detailed analysis. As we can see below, the effect in question is
possible only for the processes with asymptotically growing
cross-sections. That is why we have considered the data starting
from the ISR where $ \sigma_{tot}^{pp} $ begin to grow.

Formally, it is easy to invent a model amplitude which exhibit such
a property. Let us consider a single Regge-pole amplitude of the
form $ \beta(t)s^{\alpha (t)} $. If the Regge-trajectory $ \alpha
(t) $ intersect the line $ \alpha  = 1 $ at some negative $ t_{\ast}
$ then the differential cross-section defined by such an amplitude
is stationary at $ t = t_{\ast} $. This is possible for the Pomeron
trajectory only with $ \alpha(0)>1 $. However, there are arguments
that the Pomeron trajectory is non-linear and never intersects the
level $ \alpha  = 1 $ at negative $ t $ [4]. Sure, the genuine
scattering amplitude is much more complicated than this simplistic
model. Notwithstanding the plethora of "realistic" models which
describe the data with varied success, we estimate the differential
cross-section at the stationary point with help of a simple
quasi-classical expression
\begin{center}
$ d\sigma_{pp} = 2\pi r {dr}$.
\end{center}
Uncertainty relations imply that
\begin{center}
$  r \approx {1}/{2q}$,
$ q = \sqrt{-t} $.
\end{center}
Hence, we get the estimate of the differential cross-section at the
stationary point
\begin{center}
$ {d\sigma_{pp}}/{dt}\mid_{t =t_{\ast}} = {\pi}/{4t_{\ast}^{2}}.  $
\end{center}
After having inspected the available data we choose, as a trial
value of $ t_{\ast} $ , the value $ - 0.21$ GeV$^{2} $. Then

\begin{center}
$  ({d\sigma_{pp}}/{dt})(s,t_{\ast}) \approx 7 \mathrm{\,
mb}/\mathrm{GeV}^{2} $.
\end{center}
It is bizarre that such a primitive evaluation lies so close to the
data which we estimate at the level $\approx 7,5 \mathrm{\,
mb}/\mathrm{GeV}^{2}$ [5]-[8] (c.f. Fig.1).

\section{Estimate of the Forward Slope at 13 TeV}

Let us rewrite the differential cross-section in the form
\begin{center}
$  ({d\sigma_{pp}}/{dt})(s,t)
=({d\sigma_{pp}}/{dt})(s,t=0)\exp[\beta(s,t)] ,$
\end{center}
where \begin{center} $ \beta(s,t) \doteq \int\limits_{0}^{t}dt^{'}
B(s,t^{'}) $
\end{center}
and
\begin{center}
$ B(s,t)= d ( \ln[({d\sigma_{pp}}/{dt})(s,t)])/dt $
\end{center}
is the local slope. Existence of the stationary point means that at
$ t =t_{\ast} $, which we  take as before at $ - 0.21$ GeV$^{2}$
\begin{center}
$ ({d\sigma_{pp}}/{dt})(s,t_{\ast}) =
({d\sigma_{pp}}/{dt})(s,t=0)\exp[\beta(s,t_{\ast})] = 7.5 \mathrm{\,
mb}/\mathrm{GeV}^{2} .$
\end{center}

This quantity is, by our assumption, energy independent therefore
the growth of $ ({d\sigma_{pp}}/{dt})(s,t=0) $ is to be
compensated by the decrease of $  \beta(s,t_{\ast}) $. Taking use of
the mean value theorem we have
\begin{center}
$ \beta(s,t_{\ast}) = t_{\ast} B(s,\tilde{t}), \tilde{t} \in
[t_{\ast},0] .$
\end{center}
We obtain
\begin{center}
$ B_{pp}(s,\tilde{t})= \ln [({{d\sigma_{pp}}/{dt}|_{t=0}})/({{d\sigma_{pp}}/{dt}|_{t=t_{\ast}}})]/({-t_{\ast}}), $
\end{center}
or
\begin{center}
$  B_{pp}(s,\tilde{t})= 9.5 \ln[0.08 \,
{\sigma_{tot}^{pp}\sqrt{1+\rho^{2}}}] ,$
\end{center}
where $\rho =\mathrm{Re}T_{pp}(s,0)/\mathrm{Im}T_{pp}(s,0) $.

To estimate $ B_{pp}(s =(13 \, \mathrm{TeV})^{2},0) $ we take use of the
parametrization suggested in [9] which gives $$
\sigma_{tot}^{pp} \approx 110.0 \mathrm{\ mb} $$  at $\sqrt{s} = 13$
TeV and $\rho \approx 0.14 $.

From these values we can envisage that
\begin{center}
$ B_{pp}(s =(13 \mathrm{\ TeV})^{2},0)\geq B_{pp}(s =(13 \mathrm{\ TeV})^{2} ,
\tilde{t}) \approx 20.8 \mathrm{\ GeV}^{-2} $.
\end{center}

\section{Discussion}
We have shown, at a tentative level, that it is quite probable that
there exists a stationary, energy independent value $ t = t_{\ast}
\approx - 0.21$ GeV$^{2} $ where the value of $
({d\sigma_{pp}}/{dt})(s,t_{\ast})$  remains constant with the energy
growth. We have some reasons to believe that one another stationary
point lies near $ - 2$ GeV$^{2} $ . It seems natural to investigate
other channels : $ p\bar{p}, \pi p, Kp $ . At the moment we have no
physical explanation of the phenomenon considered above. \vskip 5mm

\section{Acknowledgements}
We are grateful to V.V.~Ezhela and A.V.~Kisselev for help and
fruitful discussions.

\end{document}